\documentclass[12pt]{article}

\usepackage{graphicx}

\begin{document}

\title{Relaxation under outflow dynamics with random sequential updating.}

\author{Sylwia Krupa and Katarzyna Sznajd-Weron \\
Institute of Theoretical Physics,\\ University of Wroc{\l}aw}

\date{\today}
\maketitle

In this paper we compare the relaxation in several versions of the Sznajd model (SM) with random sequential updating on the chain and square lattice. We start by reviewing briefly all proposed one dimensional versions of SM. Next, we compare the results obtained from Monte Carlo simulations  with the mean field results obtained by Slanina and Lavicka .
Finally, we investigate the relaxation on the square lattice and compare two generalizations of SM, one suggested by Stauffer and another by Galam. We show that there are no qualitative differences between these two approaches, although the relaxation within the Galam rule is faster than within the well known Stauffer rule.

\section{Introduction}
Applying the Ising model in the social sciences already has quite a long history \cite{g04,HKS01}.
In 2000 we proposed a simple model based on Ising spins \cite{SWS00}, which was
aimed at describing global social phenomena (sociology) by local
social interactions (described by social psychology).
The motivation to introduce the novel Ising spin dynamics was the so-called {\it Social Validation} -
one of the most powerful phenomena that influence human decision \cite{Cialdini00}. 
The fundamental way that we decide what to do in a situation is to look at what others are doing. An isolated
person does not convince others; a group of people sharing the
same opinion influences the neighbors much easier. 
 The crucial difference of our model, originally called USDF after the trade union maxim "United we Stand, Divided we Fall", compared to voter or Ising-type
models is that information flows outward. The USDF model, later renamed by Dietrich Stauffer the "Sznajd model" (SM) has been modified  and applied in marketing, finance and politics; for reviews see \cite{Stauffer,fs05,SW05Phys}.

Sociologically inspired models pose new challenges to statistical physics \cite{SL03}. Therefore, simultaneously to applications, SM has been investigated from the theoretical point of view  \cite{SO02,SL03}. Moreover, numerous modifications of SM have been proposed  \cite{Sanchez04,Ochrombel01,SSO00}. In this paper we aim at gathering the models "under a common roof". Namely, we compare the relaxation in several versions of SM with random sequential updating on the chain and square lattice.  

\section{One dimensional case}
The one-dimensional USDF model was defined in the following 
way \cite{SWS00}:
\begin{enumerate}
\item In each time step a pair of spins $S_i$ and $S_{i+1}$ is
chosen to change their nearest neighbors (nn), i.e. the spins 
$S_{i-1}$ and $S_{i+2}$.
\item If $S_iS_{i+1}=1$ then
$S_{i-1}=S_{i}$ and $S_{i+2}=S_{i}$ -- ferromagnetic rule.
\item If
$S_iS_{i+1}=-1$ then $S_{i-1}=S_{i+1}$ and $S_{i+2}=S_{i}$ -- antiferromagnetic rule. 
\end{enumerate}

Two types of attractors are possible in such a model - ferromagnetic and antiferromagnetic steady states. Both are equally probable.

Recently, attention has been paid to the second antiferromagnetic case. It was claimed
that the antiferromagnetic rule could be considered to be quite unrealistic in a model trying to represent the behavior of a community. To avoid the unrealistic 50-50 alternating final state, new dynamic
rules were proposed by Sanchez \cite{Sanchez04}:
\begin{enumerate}
\item In each time step a pair of spins $S_i$ and $S_{i+1}$ is
chosen to change their nearest neighbors (nn), i.e. the spins
$S_{i-1}$ and $S_{i+2}$.
\item If $S_iS_{i+1}=1$ then
$S_{i-1}=S_{i}$ and $S_{i+2}=S_{i}$.
\item If
$S_iS_{i+1}=-1$ then $S_{i-1}=S_{i}$ and $S_{i+2}=S_{i+1}$. 
\end{enumerate}
The model with these rules was studied by Monte Carlo simulations on a chain of length $N=100$.
It was found that the antiferromagnetic final state was completely avoided and that
the other two types of total agreement (ferromagnetic) final states could be
achieved, with equal probability, from the same initial
conditions. Moreover, the scaling exponent of the number of spins that changed their state
with time was shown to be the same like in the original Sznajd model \cite{SO02}.

This result is not very surprising, because both models (USDF \cite{SWS00} and Sanchez \cite{Sanchez04}) can be rewritten in a simpler form, namely:
\begin{itemize}
\item
USDF:
$S_{i-1}$ takes direction of $S_{i+1}$ and $S_{i+2}$ takes direction of $S_{i}$ 
\item
Sanchez:
$S_{i-1}$ takes direction of $S_{i}$ and $S_{i+2}$ takes direction of $S_{i+1}$ 
\end{itemize}
In both dynamics a spin takes the direction of another spin for any value of $S_iS_{i+1}$. 
A similar idea was introduced by Ochrombel \cite{Ochrombel01}
without qualitatively affecting many of the results (except
the presence of the phase transitions in two dimensions). In his approach a randomly chosen spin influences
its neighbors, i.e. the neighbors get the same orientation. In one dimension this rule can be written as:
\begin{itemize}
\item
Ochrombel rule:
$S_{i-1}$ and $S_{i+1}$ take direction of $S_{i}$. 
\end{itemize}

The observation that in all the described above dynamics only one spin influences other, makes  the results obtained in \cite{bs03} more clear. In \cite{bs03} it was shown that one dimensional case USDF model
could be completely reformulated in terms of a linear Voter model. 

Moreover, it is easy to notice that in the mean field approach all three models (USDF, Sanchez and Ochrombel) are equivalent. If we denote by $N_+$ the number of up spins and by $N_-$ the number of down spins then
we can write the magnetization defined as:
\begin{equation}
m=\frac{N_+-N_-}{N} = p_{+}-p_{-}=p_{+}-(1-p_{+})=2p_{+}-1,
\end{equation}
where $p_+$ is concentration of up-spins.

In one time step, three events are possible: the magnetization increases by $2/N$, 
decreases by $2/N$ or remains constant. For all three models (USDF, Sanchez and Ochrombel) these probabilities within the mean field approach for $N \rightarrow \infty$ are:

\begin{eqnarray}
\gamma^+(m) & = & {\rm Prob}\left\{ m \rightarrow m+\frac{2}{N} \right\} = p_{+}(1-p_{+}) = \frac{1-m^2}{4}, \nonumber\\
\gamma^-(m) & = & {\rm Prob}\left\{ m \rightarrow m-\frac{2}{N} \right\} = p_{+}(1-p_{+}) = \frac{1-m^2}{4}, \nonumber\\
\gamma^0(m) & = & {\rm Prob}\left\{ m \rightarrow m\right\} = 1-2p_{+}(1-p_{+}) = \frac{1+m^2}{2}.
\end{eqnarray}

Number of analytical results such as mean relaxation time or distribution of waiting times have been obtained within such an approach \cite{SL03}.

As we have already said, in all three models only one spin influences its neighbor -- the first neighbor for the Ochrombel and Sanchez modifications and the second neighbor for the USDF model. This is not very realistic from social point of view. Another possibility to avoid the antifferomagnetic final state, much more in the spirit of social validation, was briefly mentioned in \cite{SWS00} and investigated by the mean field approach in \cite{SL03}. An analogous rule was proposed by Stauffer on the square lattice \cite{SSO00}. From now on we call this rule {\it Social Validation} (SV):
\begin{enumerate}
\item In each time step a pair of spins $S_i$ and $S_{i+1}$ is
chosen to change their nearest neighbors (nn), i.e. the spins 
$S_{i-1}$ and $S_{i+2}$.
\item If $S_iS_{i+1}=1$ then
$S_{i-1}=S_{i}$ and $S_{i+2}=S_{i}$ (social validation).
\item If
$S_iS_{i+1}=-1$ then $S_{i-1}=S_{i-1}$ and $S_{i+2}=S_{i+2}$ (nothing happens). 
\end{enumerate}

This model was studied by Slanina and Lavicka \cite{SL03} using the mean field approach .
They approximated the social network by the
fully-connected network (the complete graph) of $N$ nodes.
In such a case any two agents are neighbors and this is equivalent to the mean field approach. They were able to obtain results for the mean relaxation time:  
\begin{equation}
\left< \tau_{st} \right> \approx -\ln \left( \frac{|2p-1|}{\sqrt{Np(1-p})} \right),
\end{equation} 
and deduce that the distribution of waiting times would have an
exponential tail:
\begin{equation}
P^>_{st}(\tau) \sim e^{-2\tau}, \hspace{0.5cm} \tau \rightarrow \infty,
\end{equation}
where $\tau=t/N^2=MCS/N$ is rescaled time, $t$ is an elementary time step and $MCS$ is one Monte Carlo step. 

All four presented above dynamics can be called "outflow" dynamics, since the information flows from the center spin (or spins) to the neighborhood. Of course we can investigate also "inflow dynamics" such as zero-temperature Glauber dynamics \cite{Glauber}:
\begin{enumerate}
\item In each time step a spin $S_i$ is randomly chosen.
\item If $S_{i-1}S_{i+1}=1$ then $S_{i}=S_{i+1}$ 
\item If
$S_{i-1}S_{i+1}=-1$ then $S_{i}=-S_{i}$ with probability 1/2. 
\end{enumerate}

Except for the original USDF model, all dynamics (Ochrombel, Sanchez, SV and Glauber) lead to full consensus on the chain with probability 1. A different situation is observed on the square lattice. A simple Ising ferromagnet has a large number of metastable states with respect to Glauber spin-flip dynamics \cite{SKR1,SKR2}. Therefore at zero temperature the system could get stuck forever in one of these states. There appears to be a nonzero probability that the square lattice system freezes into a stripe configuration. At $T=0$, metastable states in this dynamics have an infinite lifetime that can prevent the equilibrium ground state from being reached. 

\section{Monte Carlo results for one dimensional consensus rules}
In this section we compare the results obtained from Monte Carlo simulations on the chain of length $N=10^2$ for all proposed one dimensional consensus models, i.e. Ochrombel, Sanchez, SV and Glauber dynamics. 
Next, we compare the results obtained from Monte Carlo simulations on the chain ($N=10^2$ and $N=10^3$) with the mean field results obtained in \cite{SL03} for SV dynamics.

Probably the most natural way to investigate the relaxation process of the consensus dynamics is to look at the number of clusters in time. In all four dynamics the number of clusters monotonically decays as
$t^{-1/2}$. However, the differences between the dynamics can be observed if we look at the probability of reaching the final "all spins up" state (Fig.1) and the mean relaxation time (Fig.2) as a function of the initial fraction of up-spins $p_{+}(0)$. 

It is seen that for three dynamics (Ochrombel, Sanchez and Glauber) the dependence between the probability $P_{st}(all +)$ of the final "all spins up" state and the initial value of up-spins $p_{+}(0)$ is linear, while for SV the curve is {\it S}-shaped (see Fig.1). This reminds of the phase transition obtained in the mean field approach \cite{SL03} and for the simulations on the square lattice \cite{SSO00}.

\begin{figure}
\begin{center}
\includegraphics[scale=0.5]{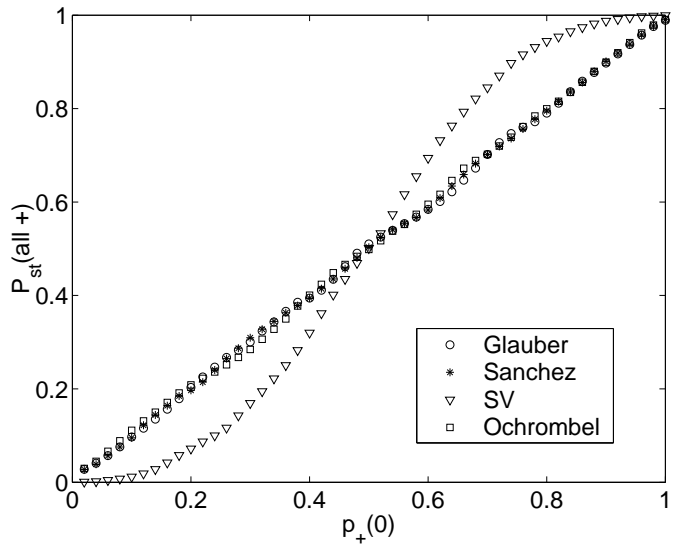}
\caption{Probabilities of reaching the final "all spins up" state from a random initial state consisting of $p_{+}(0)$ up spins for Ochrombel (squares), Sanchez (stars), SV (triangles) and Glauber (circles) dynamics.}
\end{center}
\end{figure}

\begin{figure}
\begin{center}
\includegraphics[scale=0.5]{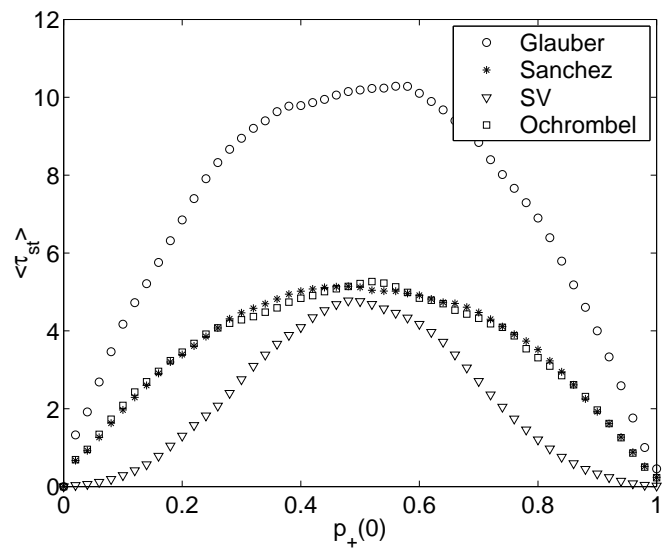}
\caption{Mean relaxation times for an $L=100$ chain from a random initial state consisting of $p_{+}(0)$ up spins for Ochrombel (squares), Sanchez (stars), SV (triangles) and Glauber (circles) dynamics.}
\end{center}
\end{figure}

The differences between the dynamics are also clear if we look at the mean relaxation time $<\tau_{st}>$ as a function of the initial value of up-spins $p_{+}(0)$. For the Ochrombel, Sanchez and Glauber dynamics this dependence is well fitted by the quadratic function, while in the case of the SV model this dependence is bell-shaped (see Fig.2). Only quantitative differences can be observed between  Ochrombel, Sanchez and Glauber dynamics - the relaxation under Glauber is much slower than under the other two.

Let us now investigate the SV case more carefully.
In Fig.3 the probabilities of reaching the final "all spins up" state as a function of the initial value of up-spins $p_{+}(0)$ for two lattice sizes ($L=10^2,10^3$) are presented. It can be observed that this dependence is not steeper for the larger lattice. Moreover, if we look at Fig.4 we see that the mean relaxation time behaves also the same for $N=10^2$ and $N=10^3$. This suggests that there is no phase transition in one dimension, in contrast to the mean field results.

\begin{figure}
\begin{center}
\includegraphics[scale=0.5]{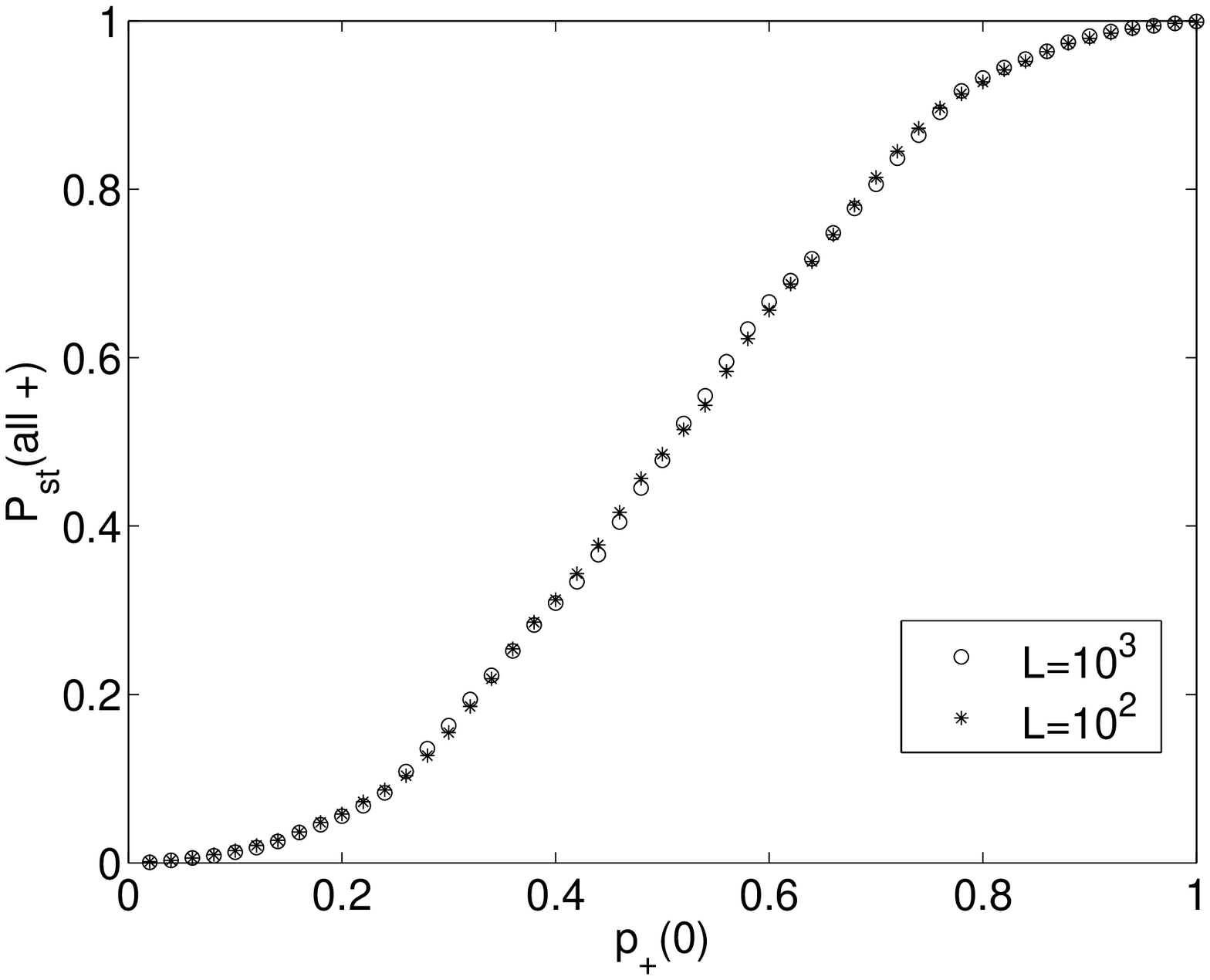}
\caption{Probabilities of reaching the final "all spins up" state from a random initial state consisting of $p_{+}(0)$ up spins for SV dynamics on the chain of length $L=10^2$ and $L=10^3$. Clearly no phase transition is present on the chain.}
\end{center}
\end{figure}

\begin{figure}
\begin{center}
\includegraphics[scale=0.5]{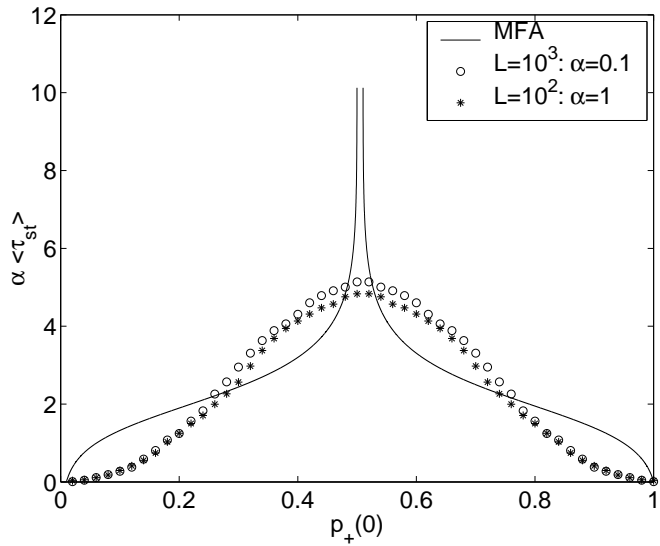}
\caption{Mean relaxation times (divided by $\alpha$, to present results for $L=10^2$ and $L=10^3$ on the same plot) from the initial state, containing randomly distributed $p_+=(m_0+1)/2$ up-spins, to the ferromagnetic steady state. Obviously the mean field results are completely different from the simulation results. Moreover, comparison of the results for $L=10^2$ and $L=10^3$ clearly shows that no phase transition is present on the chain.}
\end{center}
\end{figure}

\begin{figure}
\begin{center}
\includegraphics[scale=0.5]{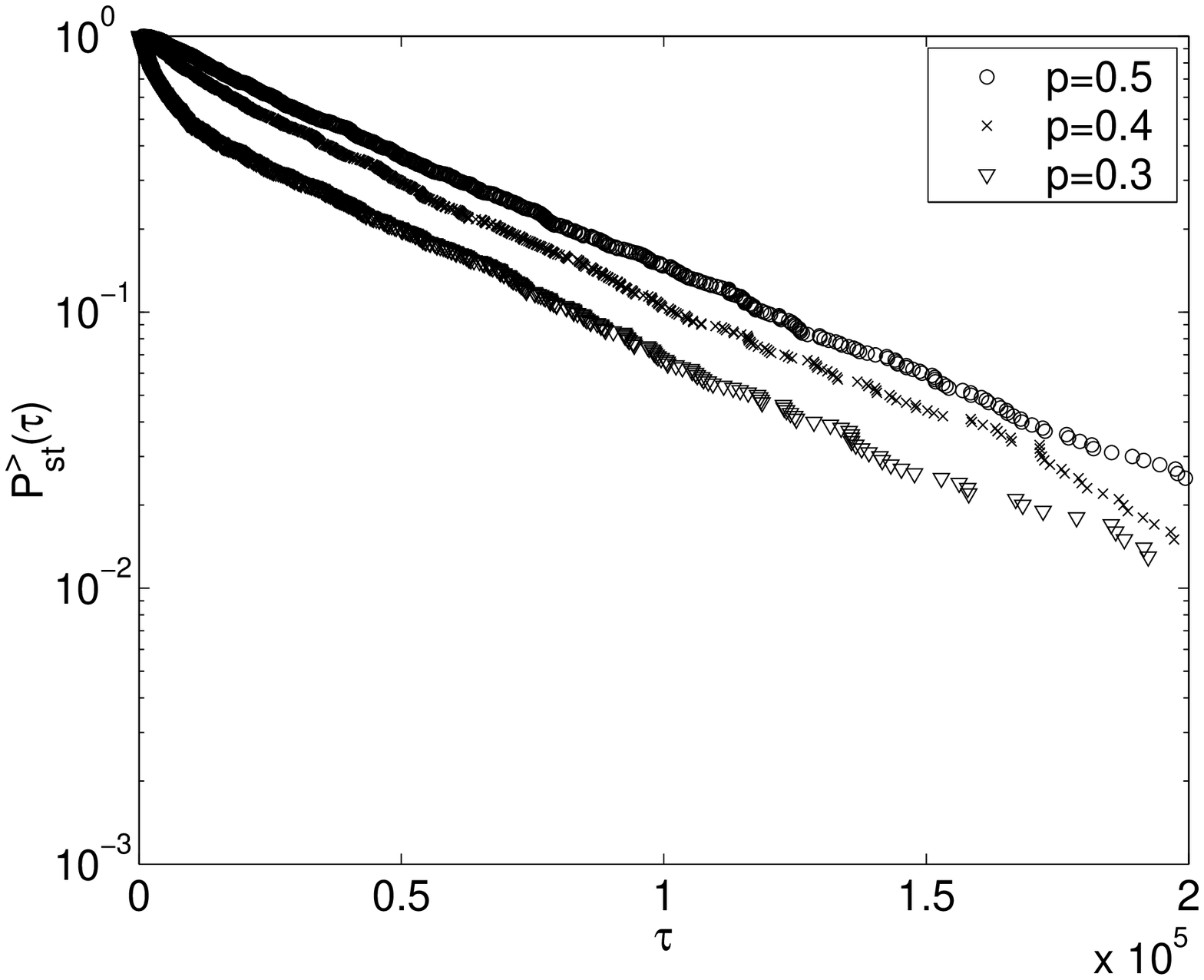}
\caption{Probabilities of reaching the steady state in time
larger than $\tau$ on the chain for several values of the cinitial fraction of up-spins $p$. 
The distribution of waiting times has an exponential tail  with the exponent independent of $p$.}
\end{center}
\end{figure}

However, the mean field results give the correct prediction on the distribution of the relaxation time. 
It was found that the distribution of waiting times had an exponential tail \cite{SL03}.
Monte Carlo simulations confirm this prediction both on the complete graph \cite{SL03} and on the chain (see Fig.5). 

\section{Two dimensions}
Two dimensional models are much more realistic than one dimensional in describing social systems. Several possibilities of generalization to the square lattice were proposed by Stauffer et al. \cite{SSO00}. They presented the model on a square $L \times L$ lattice where again every spin can be up or down. Six different rules were introduced, but only two of them have been used in later publications:
\begin{itemize}
\item A $2 \times 2$ panel of four neighbors, if not all four center spins are parallel,
leaves its eight neighbors unchanged (see Fig.1).
\item A neighboring pair persuades its six neighbors to follow the pair orientation if
and only if the two pair spins are parallel.
\end{itemize}
With both these rules complete consensus is always reached as a steady state. Moreover, a phase transition is observed - initial densities below 1/2 of up-spins lead to all spins down and densities above 1/2 to all spins up for large enough systems \cite{SSO00}.
Galam (priv. comm. with Stauffer, described in \cite{SSO00}) showed that the updating rule of the one-dimensional SM can be transformed exactly into two dimensions in the following way (see Fig.1): the one-dimensional rule is applied to each of the four chains of four spins each, centered about two horizontal and two vertical pairs.   

\begin{figure}
\begin{center}
\includegraphics[scale=0.5]{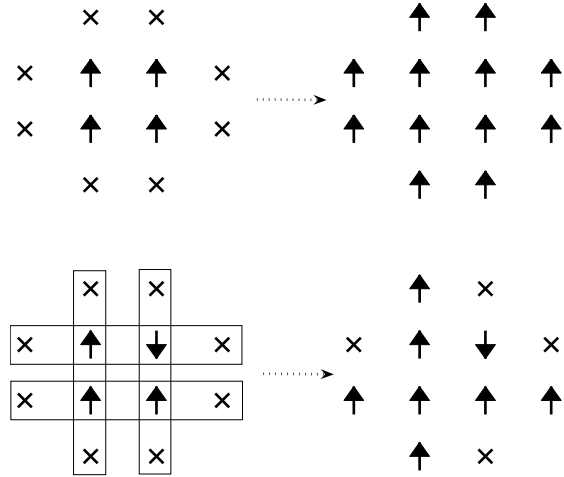}
\caption{Dynamical rules of SM on the square lattice proposed by Stauffer  (top) and Galam (bottom). 
In every time step a panel of four spins denoted by arrows is picked at random and influences spins denoted by $x$ producing the configurations presented in the right-hand side pictures.}
\end{center}
\end{figure}

Here we compare two rules in which a panel of four spins influences eight neighbors, i.e. Galam (Galam dynamics) and first of Stauffer's rules (Stauffer dynamics). It seems that Stauffer dynamics is more attractive from the social point of view \cite{AWA94}.
Salomon Asch, in his famous experiment on conformity, found that one of the situational factors that influence conformity is the size of the opposing majority. In a series of studies he varied the number of confederates who gave incorrect answers from 1 to 15. He found that the subjects conformed to a group of 4 as readily as they did to a larger group. However, the subjects conformed much less if they had an "ally". Apparently, it is difficult to be a minority of one but not so difficult to be part of a minority of two. 

On the other hand Galam dynamics is very simple for generalization on any regular lattice and was used   to construct a two dimensional version of the so called two-component model \cite{SW04}.

We have measured the probability of reaching the final "all spins up" state, as well as the mean relaxation time from a random initial state consisting of $p_{+}(0)$ up spins for Stauffer and Galam dynamics. We have found the phase transition for both dynamics - for $p_{+}(0)<0.5$ the "all spins up" state is never reached, while for $p_{+}(0)>0.5$ this state is obtained with probability 1 (this result was obtained in \cite{SSO00}). Moreover, critical slowing down is observed at $p_{+}(0)=0.5$ (see Fig.7). For $L \rightarrow \infty$ we expect the $\delta(0.5)$ - function.

\begin{figure}
\includegraphics[scale=0.7]{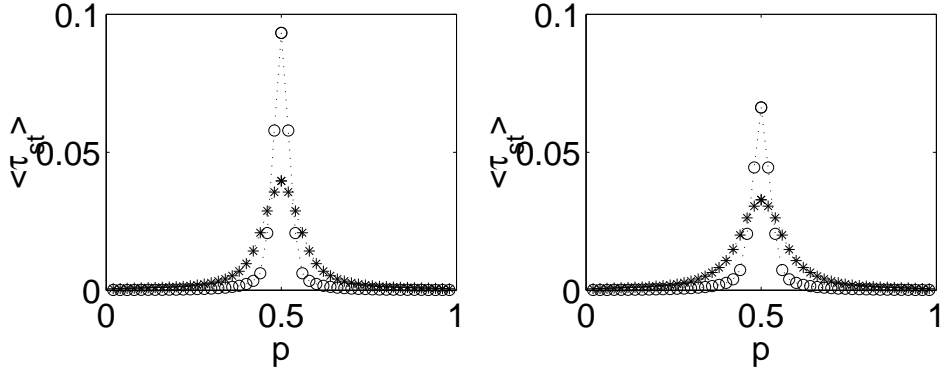}
\caption{Mean relaxation times from the initial state, containing randomly distributed $p_+(0)=(m_0+1)/2$ up-spins, to the ferromagnetic steady state for Stauffer (left panel) and Galam (right panel) dynamics. 
Results are presented for the square lattice $30 \times 30$ (*) and $100 \times 100$ (o). The phase transition is clearly seen for both dynamics at $p_+(0)=0.5$. However, the relaxation under Glauber dynamics is faster than under Stauffer dynamics.}
\end{figure}

We have also investigated the distribution of the relaxation time. In the mean field approach \cite{SL03} and on the chain (previous section) we have found that the distribution of waiting times has an exponential tail with 
$p$-independent exponent ($p=p_+(0)$). Results for the square lattice are presented in Fig.8. Under both dynamics the distribution of relaxation times has an exponential tail, but the exponent is $p$-dependent. Interestingly, the dependence between the exponent and the initial number of up-spins is identical for both dynamics.
 
\begin{figure}
\includegraphics[scale=0.7]{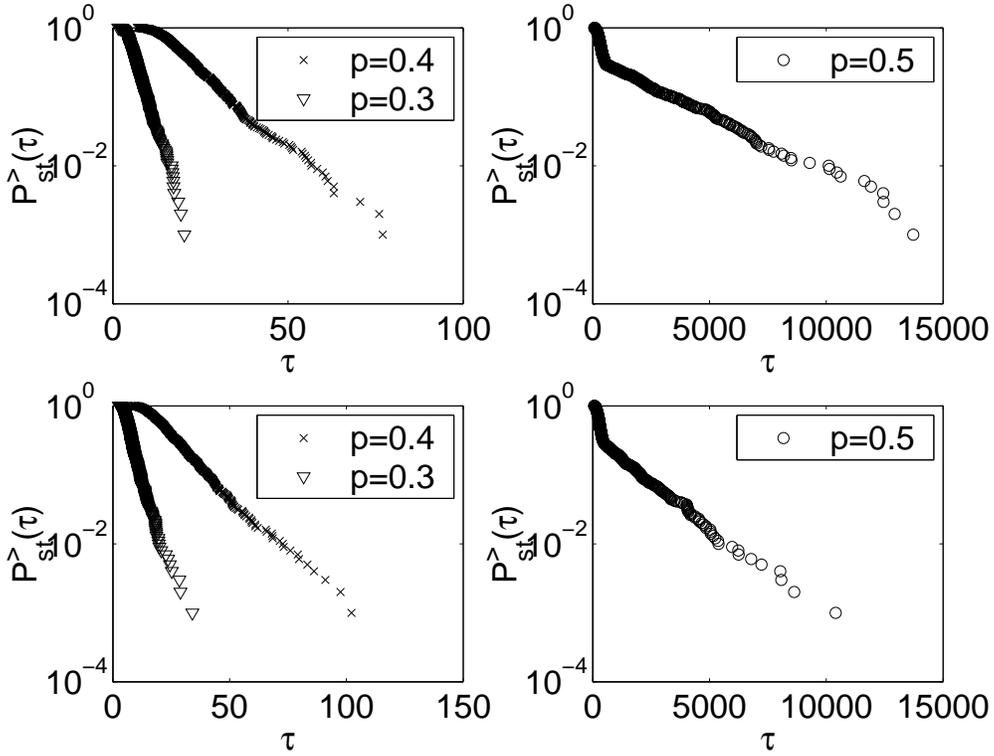}
\caption{Probabilities of reaching the steady state in time
larger than $\tau$ on the square lattice for several values of the initial fraction of up-spins $p$ for Stauffer (top) and Galam (bottom) generalizations of SM.  
The distribution of waiting times has an exponential tail  with the exponent dependent on $p$. For both dynamics this dependence is identical. Critical slowing down is observed for $p=0.5$ (right panels).}
\end{figure}

Finally, we have measured the number of clusters $N_c$ in time under both dynamics. We have found that for the initial fraction of up-spins $p=0.5$ the number of clusters  decays in time as a power law $N_c \sim t^{-0.5}$ (see Fig.9), which is the same result as for the chain. Again, for any initial value of $p$ the relaxation under Galam and Stauffer dynamics are quantitatively the same.

\begin{figure}
\begin{center}
\includegraphics[scale=0.5]{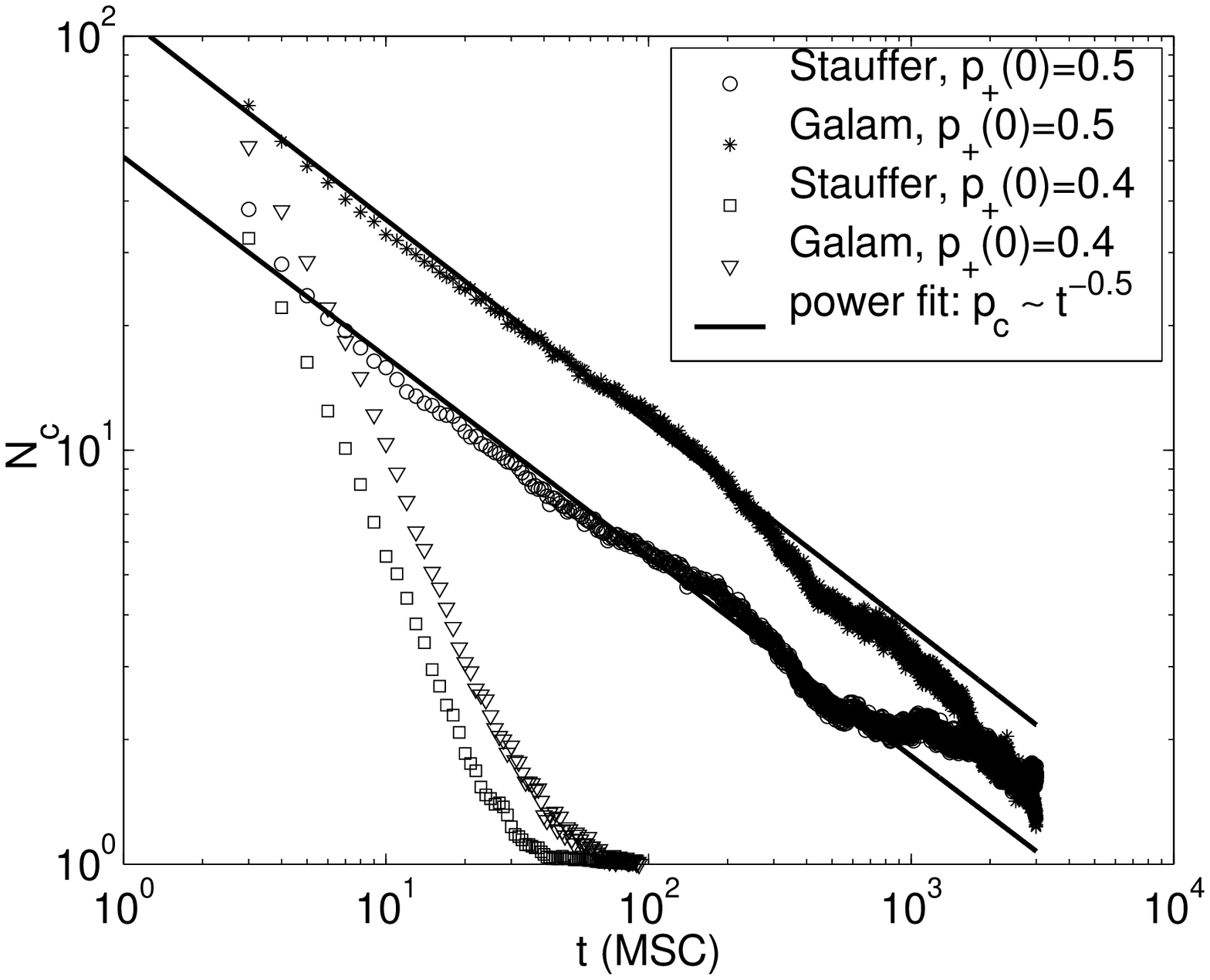}
\caption{The number of clusters decays as a power law for both dynamics.}
\end{center}
\end{figure}

\section{Conclusions} 
We have compared several one and two dimensional versions of SM. We have investigated the probability and the mean relaxation time of reaching the final "all spins up" state  
from the initial state, containing randomly distributed $p_+(0)$ up-spins. Moreover, we have measured the distribution of relaxation times and the decay of the number of clusters. 
We have found that on the chain Sanchez and Ochrombel modifications give qualitatively the same results as the zero-temperature Glauber dynamics, although time needed to reach the steady state under Glauber dynamics is much longer. In contrast the one dimensional dynamics proposed by Slanina and Lavicka in \cite{SL03}, which we have called Social Validation dynamics is completely different. The difference is visible both in the probability and the mean relaxation time of reaching the final "all spins up" state   
from the initial state, containing randomly distributed $p_+(0)$ up-spins. For SV dynamics also analytical results were obtained \cite{SL03} and we were able to compare them with our simulation results. As often happens the results were remarkably different. In particular, no phase transition predicted by the mean field theory is observed in the one dimensional system. 

In the second part of this paper we have compared the results given by Stauffer and Galam generalizations of SM. This comparison seems to be quite important, since Stauffer generalization seems to be more attractive from social point of view, while Galam rule is much easier for generalization to other systems (in particular, it was used in the so called TC model \cite{SW04}). No qualitative difference has been found between these two dynamics, although time needed to reach the final state is shorter under Galam dynamics.

\end{document}